\documentclass[superscriptaddress,twocolumn]{revtex4}
\usepackage{graphicx,color,url,ulem}
\usepackage{amsmath}
\usepackage{amssymb}
\usepackage{dcolumn}

\definecolor{darkgreen}{RGB}{0,139,0}
\definecolor{turqoise}{RGB}{64,224,208}

\definecolor{b}{rgb}{0,0,1.0}
\definecolor{r}{rgb}{1,0,0}
\definecolor{g}{rgb}{0,1,0}
\definecolor{rs}{rgb}{0.5,0.0,1}

\def\mueff{\mu_{\text{eff}}}

\begin{document}

\newcommand{\OK}[1]{{\color{darkgreen}{#1}}}
\newcommand{\Quest}[1]{{\color{r}({#1})}}
\newcommand{\MOD}[1]{{\color{b}{#1}}}
\newcommand{\RS}[1]{{\color{rs}{#1}}}

\newcommand{\SZFKI}{Institute for Solid State Physics and Optics, Wigner
Research Center for Physics, Hungarian Academy of Sciences, P.O. Box 49, H-1525
Budapest, Hungary}

\newcommand{\BME}{Institute of Physics, Budapest University of Technology
and Economics, H-1111 Budapest, Hungary}

\newcommand{\MDB}{Otto-von-Guericke-University, D-39106 Magdeburg, Germany}

\newcommand{\MDBL}{Leibniz Institute for Neurobiology, D-39118 Magdeburg, Germany}

\title{Evolution of shear zones in granular materials}

\author{Bal\'azs Szab\'o}

\email{szabo.balazs@wigner.mta.hu}
\affiliation{\SZFKI}
\author{J\'anos T\"or\"ok}
\affiliation{\BME}
\author{Ell\'ak Somfai}
\affiliation{\SZFKI}
\author{Sandra Wegner}
\affiliation{\MDB}
\author{Ralf Stannarius}
\affiliation{\MDB}
\author{Axel B\"ose}
\affiliation{\MDB}
\author{Georg Rose}
\affiliation{\MDB}
\author{Frank Angenstein}
\affiliation{\MDBL}
\author{Tam\'as B\"orzs\"onyi}
\affiliation{\SZFKI}

\begin{abstract}
The evolution of wide shear zones (or shear bands) was investigated experimentally and numerically for
quasistatic dry granular flows in split bottom shear cells.
We compare the behavior of materials consisting of beads, irregular grains (e.g. sand) and elongated particles.
Shearing an initially random sample, the zone width was found to significantly decrease in the first stage
of the process. The characteristic shear strain associated with this decrease is about unity and it
is systematically increasing with shape anisotropy, i.e. when the grain shape changes from spherical to
irregular (e.g. sand) and becomes elongated (pegs).
The strongly decreasing tendency of the zone width is followed by a slight increase which is more pronounced
for rod like particles than for grains with smaller shape anisotropy (beads or irregular particles).
The evolution of the zone width is connected to shear induced density change and for nonspherical particles
it also involves grain reorientation effects.
The final zone width is significantly smaller for irregular grains than for spherical beads.
\end{abstract}

\maketitle

\section{Introduction}
\label{ff1}

Shear banding is often observed in granular materials and is an important factor in various industrial
and geological processes.
For dry granular materials several features of the stationary flow have been explored in the last two decades
for narrow or wider shear bands. One convenient model system to investigate shear localization is the flow of beads
near the moving boundary in a cylindrical Couette geometry \cite{losert2000,veje1999}. Here the mean
velocity was found to decay rapidly with distance from the moving surface \cite{losert2000}.
Recent discrete element simulations of simple plane shear show, that shear localization near smooth walls
is even more enhanced \cite{shojaaee2012}.
When dealing with nearly uniform grains, magnetic resonance imaging (MRI) and x-ray computed tomography 
(x-ray CT) gave quantitative
information about the shear profile inside the granular material in the stationary state, showing that uniform
grain shape leads to strong layering in shear flow \cite{mueth2000}. Moreover, spherical beads with
small size dispersion were shown to crystallize due to shear \cite{tsai2004,wegner2014}, leading to compaction and decreased
shear stress.

The case of wider shear bands is even more puzzling since their width results from the complex rheology of
the granular flow including various factors.  The so called split bottom geometry \cite{dijksman2010} is
suitable to investigate these wider shear bands which develop in the bulk (away from the boundary).
Experimentally, the cylindrical configuration is the most convenient to realize stationary
shear rates and quantify the position and width of the shear band, which has been done for a variety of materials
\cite{fenistein2003,fenistein2004,fenistein2006,cheng2006}.
To explain why these zones are wider than expected, the most important ingredients are probably nonlocal aspects
of the rheology \cite{nichol2012,wortel2014,henann2013}, i.e. flow induces agitations in the neighborhood and
thereby contributes to yielding of the material.
Recently, a continuum model based on this approach was very successful to reproduce the stationary sizes
of the shear zones in various geometries \cite{henann2013}. Similar nonlocal effects had to be
incorporated in earlier numerical simulations using the so called fluctuating narrow band model
\cite{unger2004,torok2007,moosavi2013} in order to reproduce the experimentally observed features.
An other important ingredient could be the competition between the organizational tendencies of the flow and
the gravitational pull \cite{depken2006}. The shear flow leads to the dilation of the material with respect to
the random close-packed state. The number of grain contacts is increased in the compressional direction
and decreased in expanding directions, while gravitational forces locally might favor a different anisotropy
of the contact numbers. According to Depken et al.~\cite{depken2006,depken2007} this might easily lead to a
nonuniform effective friction across the zone (having minimum friction in the middle of the zone), which would
contribute to the widening of the zone. MRI experiments in this system provided information about the Reynolds
dilation of the material under continuous shear \cite{sakaie2008}, and showed that dilatancy grows with
accumulated strain $\gamma$ and saturates when the local strain reaches the order of one. 
The fully dilated stationary state obtained at large strain is often referred to as ``critical state'' 
in the engineering literature. In that state, the rheological properties do not change any more.

Much less is known about the evolution of the system before the shear band finds the final
configuration, starting from a random configuration or reversing the shear direction.

This complex question was mostly studied {\it numerically} for spherical or nearly spherical grains.
Soft particle discrete element method (DEM) simulations with spheres provided information about the evolution
of the stress intensity, and indicated that a local strain of about $\gamma\approx 2$ is needed to
reach the stationary stress level \cite{luding2008}.
DEM simulations by Ries et al.~\cite{ries2007} showed that the zone shrinks when approaching the
stationary state, which was attributed to the the fact that time needed to locally reach the above mentioned
critical state depends on the spatial position. More explicitely, during shearing the material undergoes
shear hardening where new contacts are created against the shear. The typical strain scale for this
process was estimated to be $\gamma=0.2$, which is reached earlier in the center of the zone than in the
outer regions. In the early stage of the process when the critical zone is smaller than the shear zone,
the shear rate is slightly enhanced outside the critical region, leading to a wider zone.
After the whole zone reaches the critical state, this additional widening effect ceases and the zone
converges to its stationary asymptotic width.
Very recent numerical DEM simulations by Az\'ema et al.~investigate the time evolution of the density of a
sheared system consisting of nonspherical particles (rigid aggregates of overlapping spheres).
The scenario is very similar for various particle shapes: a quick density decrease is found which
saturates after a cumulative strain of about 0.5  \cite{azema2013}.

{\it Experimentally,} a layer of photoelastic disks exposed to shear in a 2D Couette system was investigated
by Utter and Behringer, where the system relaxed faster for the case of initially uniformly distributed particles 
than for reversed shear \cite{utter2004}. In that paper, the rearrangement of the force network associated with 
the transient was investigated. More recent 2D experiments focused on the effect of particle shape using
disks or ellipses \cite{farhadi2014}, where the shear induced rotation of the ellipses resulted in  
richer dynamics (compared to the case of disks) as well as complex density profiles.
In three dimensional systems the evolution of the packing density under shear was sucessfully determined using 
x-ray imaging for materials with various particle shapes \cite{wegner2014,kabla2009}.
Further experiments focusing on the case of reversed shear with glass beads in a 
Taylor-Couette cell \cite{toiya2004} showed, that during the transient the system compacts, the shear force is small 
and the shear band is wide, corresponding to the rearrangement (reversal) of the force network over a characteristic 
strain of $\gamma= 0.5$.

In this paper, we aim to fill the gap by presenting quantitative experimental results on the 
evolution of a three-dimensional (3D) system. We compare the behavior of materials consisting of beads, 
irregular grains (e.g. sand) and elongated particles, focusing on
changes of zone width, packing density of the material, and orientational ordering for the case of elongated grains.
We study two protocols: (i) the system is started from a random initial configuration and (ii) the shearing
direction is reversed after a stationary configuration had been achieved. We also present a numerical model, 
which is capable to reproduce several features (e.g. the non-monotonic change of the zone width) of the transient.

\section{Experimental methods and materials used}
\label{exp}

\begin{figure}[!ht]
\includegraphics[width=\columnwidth]{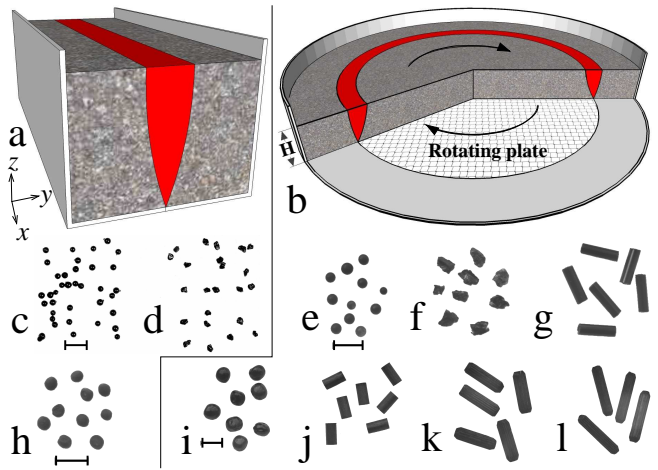}
\caption{(color online). Schematic view of the two experimental geometries used: (a) straight and (b) circular split-bottom cell.
(c)-(l) Photographs of the materials used: (c) spherical glass beads, (d) sand particles, (e) spherical
silica gel beads, (f) irregular shaped corundum particles, (g) glass rods, (h) rape seed, (i) peas, and
(j-l) wooden pegs with three different elongation. The bars correspond to (c-d) $1$ mm, (e-h) $5$ mm, and (i-l) $10$ mm.
The grain parameters are indicated in Table \ref{table-mat}.
}
  \label{setup}
\end{figure}

The experiments were carried out in two geometries: (i) straight split bottom cell and (ii) cylindrical
split bottom cell as shown in Figs.~\ref{setup}(a) and \ref{setup}(b), respectively. In the first setup, the
granular sample with height $H$ and width $4.0$ cm is sheared by displacing the two L-shaped sliders parallel
to each other.
%
%
In the second setup, shearing is realized by rotating the plate which is placed below the
material at the bottom of the container. Here $H$ is measured above the rotating plate.
Two containers were used with radius of $12$ cm and $28.5$ cm and the corresponding rotating plates had a radius
of $R=7.5$ cm and $19.5$ cm, respectively.
In the experiments various materials were used, photographs are shown in Figs.~\ref{setup}(c-l), and the main
parameters are given in Table \ref{table-mat}.
\begin{table}[!ht]
\begin{ruledtabular}
\begin{tabular}{ldddl}
\multicolumn{1}{l}{material} &
\multicolumn{1}{c}{$d$ (mm)} &
\multicolumn{1}{c}{$L$ (mm)} &
\multicolumn{1}{c}{$L/d$}    &
\multicolumn{1}{l}{image}  \\
\hline
\hline
glass beads       & 0.25   &       &      &Fig.~\ref{setup}(c)\\[-0.3mm]
sand              & 0.25   &       &      &Fig.~\ref{setup}(d)\\[-0.3mm]
\hline
silica gel beads  & 1.8    &       &      &Fig.~\ref{setup}(e)\\[-0.3mm]
corundum          & 1.8    &       &      &Fig.~\ref{setup}(f)\\[-0.3mm]
glass rods        & 1.9    & 6.6   & 3.5  &Fig.~\ref{setup}(g)\\[-0.3mm]
\hline
rape seeds        & 1.8    &       &      &Fig.~\ref{setup}(h)\\[-0.3mm]
\hline
peas              & 7.1    &       &      &Fig.~\ref{setup}(i)\\[-0.3mm]
pegs              & 5      & 10    & 2    &Fig.~\ref{setup}(j)\\[-0.3mm]
pegs              & 6      & 20    & 3.3  &Fig.~\ref{setup}(k)\\[-0.3mm]
pegs              & 5      & 25    & 5    &Fig.~\ref{setup}(l)\\[-0.3mm]
\end{tabular}
\caption{Main particle parameters: diameter $d$, length $L$ and aspect ratio $L/d$ of the materials used.}
\label{table-mat}
\end{ruledtabular}
\end{table}
The grain diameter of the irregular grains (sand and corundum) was adjusted by sieving the material, so that
the equivalent grain diameter (defined as the diameter of a sphere with the same volume) matches the diameter
of the corresponding spherical beads (glass and silica gel).
Two kinds of experiments were carried out: (i) optical detection of the distortions at the surface of the materials and
(ii) tomographic (MRI and X-ray CT) detection of the internal distortion of the sample during shear.
The experimental methods are summarized in Table \ref{table-exp}, where we have also listed the figures
in which the corresponding results are presented, including those obtained by numerical simulations.
Each experiment will be described in detail in the corresponding section.
\begin{table}[!ht]
\begin{ruledtabular}
\begin{tabular}{l|c|c}
measurement \hspace*{0.0cm}    & \hspace*{0.1cm} straight cell \hspace*{0.1cm}           & \hspace*{0.0cm} cylindrical cell \hspace*{0.0cm}    \\[-0.0mm]  \hline \hline
optical (surface)              & Figs.~\ref{width-illustration},\ref{time-width-glass}   & Figs.~\ref{time-width-circular},\ref{CT-2}          \\[-0.3mm]  \hline
MRI (bulk)                     & Figs.~\ref{width-depth},\ref{width-density},\ref{CT-2}  &                                                     \\[-0.3mm]  \hline
CT (bulk)                      &                                                         & Figs.~\ref{CT-1},\ref{CT-2}                         \\[-0.3mm]  \hline
model (bulk, cross section)                   & \multicolumn{2}{c}{Figs.~\ref{time-width-glass},\ref{time-width-circular},\ref{Fig:size},\ref{Fig:sigmad},\ref{Fig:change_d},\ref{Fig:rho_t}}    \\[-0.3mm]
\end{tabular}
\caption{Summary of the experimental methods.}
\label{table-exp}
\end{ruledtabular}
\end{table}
%

\section{Experimental results}
\label{res}
\subsection{Surface measurements}
\label{res1}

\begin{figure}[!ht]
\includegraphics[width=\columnwidth]{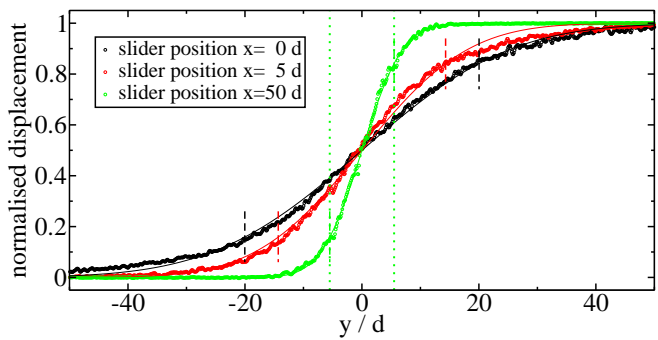}
\caption{(color online). Normalized grain displacement profiles observed at the surface of the sample in the straight shear cell at three
stages of the process, at total slider displacements of  0, $5 d$ and $50 d$, resp.
Each profile is obtained by a small displacement $dx$ of one L shaped slider, and normalized by $dx$.
Two datasets correspond to the transient, while the third one describes
the stationary state. The datasets were fitted with Eq.~\ref{erf} (thin solid lines). The fits yielded the
zone widths $w$, which are marked with dashed lines for all three cases.
Experiments were done with spherical glass beads of $d=0.25$ mm starting with a random initial configuration
setup and filling height $H=68 d$.
}
  \label{width-illustration}
\end{figure}

First we demonstrate the basic features of the evolution of the system by experimental data taken in the {\it straight
cell}. Three displacement profiles obtained by our PIV algorithm at the surface of the sample
are shown in Fig.~\ref{width-illustration}.

%
%

Two datasets correspond to the transient, while the third one was taken during the stationary state.
As seen, the shear zone is considerably wider at the beginning of the process (slider displacements 
$x=0d$ and $x=5d$) when compared to the stationary state ($x=50d$). To quantify the width of the zone 
we fit the normalized displacement data by the following function:

\begin{equation}
D(y)= \frac{1}{2} \left(1+\text{erf}\;\frac{y-y_{0}}{\sqrt{2}\, w}\right),
\label{erf}
\end{equation}
where $y$ is the coordinate along the direction of the shear gradient.

\begin{figure}[!ht]
\includegraphics[width=\columnwidth]{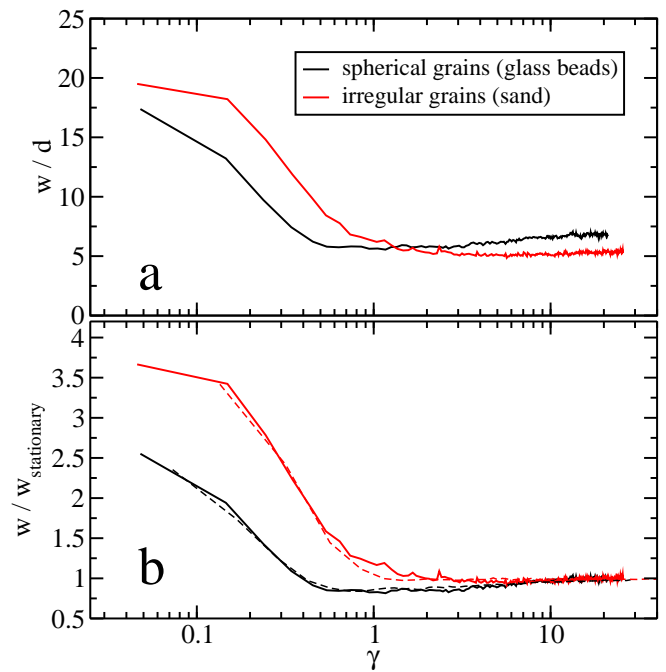}
\caption{(color online). (a) The zone width $w$ and (b) the normalized zone width $w/w_{\rm stationary}$
as a function of strain $\gamma$ in the straight cell. Experiments were done with a randomly
packed initial configuration both for spherical beads
of $d=0.25$ mm and sand particles with equivalent diameter $d=0.25$ mm.
Each curve corresponds to the average of three measurements with filling height of $H=68 d$.
The dashed lines are obtained by the numerical model, as described in Sec. \ref{numres}.
}
  \label{time-width-glass}
\end{figure}

The resulting zone width $w$ is then calculated by the above procedure at every shear step during the experiment.
We also calculate the local deformation of the material at every instant and define the strain $\gamma$
by averaging the local strain within the region corresponding to the final (stationary) zone width, i.e.
between the dotted lines in Fig.~\ref{width-illustration}.
Finally, to quantify the evolution of the zone width, $w$ is plotted as a function of $\gamma$.
These curves are shown in Fig.~\ref{time-width-glass} for glass beads and sand particles with $d=0.25$ mm,
obtained in the straight cell. Each curve is the average of three measurements.

%
%

\begin{figure*}[!ht]
\includegraphics[width=\textwidth]{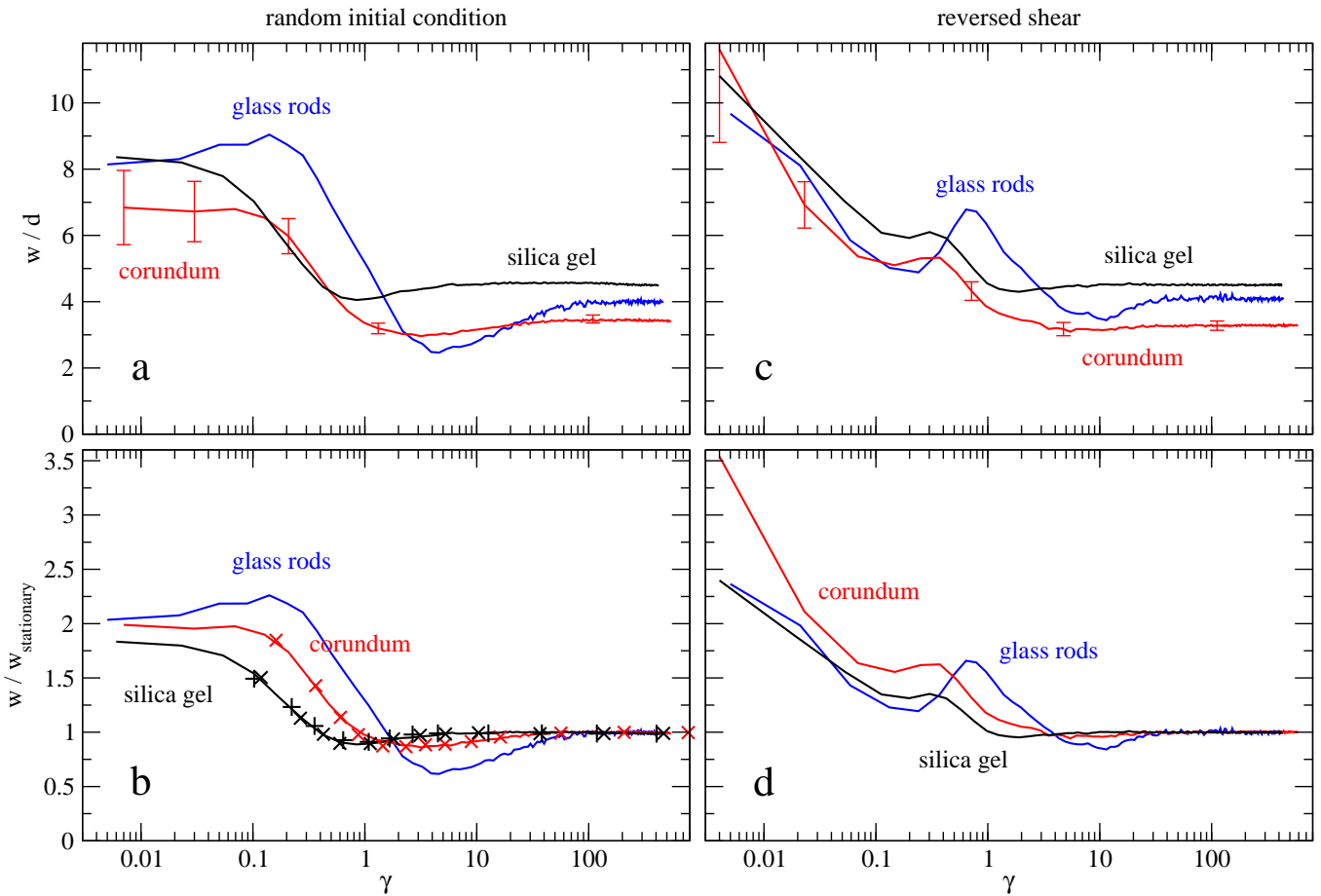}
\caption{(color online). The zone width as a function of strain $\gamma$ for three materials,
with two initial conditions. The continuous lines correspond to experiments performed
in the cylindrical cell. Each curve is the average of 100 measurements with filling height
of $H=28$ mm, corresponding to $H/d=15.6$ for silica gel beads and corundum and
$H/d=14.7$ for the glass rods. The experimental data were fitted with results of
the fluctuating narrow band model using refresh probabilities defined in Eq.~(\ref{Eq:P}) ($+$)
and Eq.~(\ref{Eq:Ptau})  ($\times$).
}
  \label{time-width-circular}
\end{figure*}

Similar measurements have been carried out in the {\it cylindrical cell}. This geometry has a strong advantage over
the straight cell, as there is no limitation for the shear displacement, thus it is easier to do longer measurements.
As a consequence, one can also simply test the system's response to reversing the direction of shearing.
In Fig.~\ref{time-width-circular} we directly compare the system's evolution starting from a
random configuration (a-b), and after reversing the shear direction (c-d) for the case of silica gel beads,
corundum particles and glass rods with L/d=3.5. Our random configuration was generated by pouring the material
into the container. In this cylindrical geometry each curve was obtained by averaging 100 measurements,
therefore the resulting evolution of $w$ is even more accurate compared to the data obtained with the
straight cell.

%
%
%

\subsubsection{Random initial configuration}

Let us first focus on the case of initially random configuration (Figs.~\ref{time-width-glass} and
\ref{time-width-circular}(a-b)). Generally, we observe that the zone width $w$ considerably decreases
as a result of the transient. This process starts when the strain reached about $\gamma\approx 0.1$
and stops when $\gamma$ reaches unity. This stage of the evolution corresponds to the formation of new contacts
between the grains and the development of an anisotropic force network from the initially random state.
Looking at the details of the data obtained with {\it spherical particles} (glass and silica gel beads) and
{\it irregular grains} (sand particles and corundum) the following observations should be noted:

(1)
The decrease of the shear zone width $w$ is more pronounced for irregular grains than for spheres. 
The ratio of the initial and stationary zone width was about 1.4 times larger for sand than for 
spherical glass beads at a filling height of $H/d = 68$ (see Fig.~\ref{time-width-glass}(b)), 
while this ratio was 1.1 when comparing the cases of corundum and silica
gel beads at a filling height of $H/d=15.6$  (see Fig.~\ref{time-width-circular}(b)).

(2)
The strain $\gamma$ needed for contraction of the shear zone width $w$ is larger 
for irregular grains. The zone width starts decreasing when the amplitude of the strain 
$\gamma\approx 0.1$ and stops decreasing around $\gamma$ reaches unity.
These numbers are approximately a factor of 2 smaller for the case of spherical beads compared
to irregular grains (see Fig.~\ref{time-width-circular}(b)).

(3) The stationary value of $w$ (in grain diameter units) is about 30 $\%$ larger for the case 
of spherical beads (glass or silica gel) than for irregular particles (sand or corundum) 
(see Figs.~\ref{time-width-glass}(a) and \ref{time-width-circular}(a)).

(4) The rapid decrease of $w$ is followed by a slight increase on a longer time scale.

The case of {\it elongated particles} with aspect ratio of $L/d=3.5$ (glass rods) is somewhat more complex, as it
involves the development of orientational ordering of the grains due to the shear flow
(\cite{borzsonyi2012,wegner2012,borzsonyi2012-2}). This has several effects on the evolution of the zone width $w$:

(1) The zone width $w$ slightly increases before it starts to decrease rapidly.
This slight increase in $w$ can be connected to the apparent lateral extent of the particles.

(2) The shear strain needed for the decrease of $w$ is significantly (3-6 times) larger than for the other
two materials (corundum and silica gel beads).

(3) The stationary value of $w$ (measured in units of d) is in-between the two values measured
for spheres and irregular grains (Fig.~\ref{time-width-circular}(a)).

(4) The rapid decrease of $w$ is followed by a significant increase on a longer time scale.

\subsubsection{Reversing the shear direction}

The initial condition for these experiments is prepared by continuous shear. Starting from the 
asymptotic configuration, opposite shear is applied and the subsequent evolution of the system 
is recorded. As seen in Figs.~\ref{time-width-circular}(c-d), reversing the shear direction 
leads to an instantaneous increase of the zone width for all three materials. Interestingly, 
$w$ jumps up to a value which is larger than its initial value for the random starting condition.
This is connected to the fact, that the particle contacts and the force network built up during 
forward shear weaken immediately after the shear direction is reversed.  In other words, the 
particles can easily move backwards into the voids which were created behind them during shearing.
This process is finished by the time the strain reaches $\gamma\approx 0.1$. At this point the zone 
width stops decreasing (for the case of glass rods it even increases) until $\gamma$ reaches about $0.5$. 
This stage of the evolution is probably connected to the establishment of a new force network, 
where the dominant contacts will be approximately perpendicular to the former ones \cite{utter2004}. 
In the final stage, $w$ continues shrinking until it reaches its stationary value.

\subsection{Bulk measurements}
\label{res2}

\subsubsection{MRI measurements in the straight cell}

In the MR tomography experiments, the whole cell was placed into a Bruker BioSpec 47/20 Magnetic Resonance Imaging (MRI)
scanner operating at 200 MHz proton resonance frequency (4.7 T) at the Leibniz Institute for Neurobiology, Magdeburg.
The cross section of our experimental apparatus was optimized for the geometry of this device with $7$ cm internal
coil diameter. The best contrast is obtained with seeds containing oil, therefore we used rape seeds
%
%
\begin{figure}[!ht]
\includegraphics[width=\columnwidth]{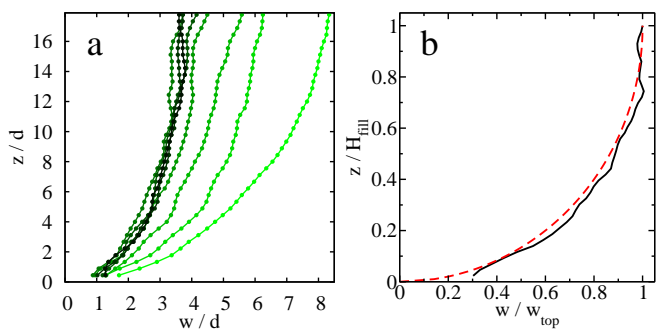}
\caption{(color online). (a) The zone profile in the cross section at different total displacements, for rape seed particles obtained
with MRI using the straight cell.
The actual total strain for the curves from green to black (right to left) 
are $\gamma=0.01$, $0.07$, $0.17$, $0.3$,
$0.47$, $0.68$, $1.34$, $2.06$, respectively.
(b) The normalized zone profile in the stationary state, and the expected curve according to Ries et al. \cite{ries2007}.
}
  \label{width-depth}
\end{figure}
with diameter $d=1.8$ mm. Horizontal slices were obtained with interslice distances of $0.8$ mm
and an in plane resolution of $0.156$ mm/pixel. The slider was displaced by $1.92$ mm (equivalent of $1.07 d$)
between subsequent MRI scans. The experiment presented here involved 25 displacement steps, yielding a total
displacement of $4.7$~cm ($26 d$). The total measurement took about 4 hours.
The displacement profile was measured in each step for each horizontal slice, similarly to the optical 
measurements presented above, and the zone width $w$ was determined by the same approach as presented in 
Fig.~\ref{width-illustration}.

The evolution of the zone width $w$ inside the sample, as detected by this procedure, is shown in
Fig.~\ref{width-depth}(a). In accordance with the data obtained for silica gel beads with similar size and filling
height (Fig.~\ref{time-width-circular}(a-b)), we find for the nearly spherical rape seeds that the zone width
considerably decreased and reached a stationary value. When we compare the two cases quantitatively, we find
the same characteristic strain scales, i.e. the stationary state is reached at $\gamma \approx 0.5$.

For the stationary state, numerical simulations predicted that the normalized zone profile
$z(w/w_{\rm top})/H_{\rm fill}$ is a quarter of a circle \cite{ries2007,singh2014}. Fig.~\ref{width-depth}(b) shows, 
that the experimentally obtained curve (solid line) nicely matches the expected form (dashed line).

\begin{figure}[!ht]
\includegraphics[width=\columnwidth]{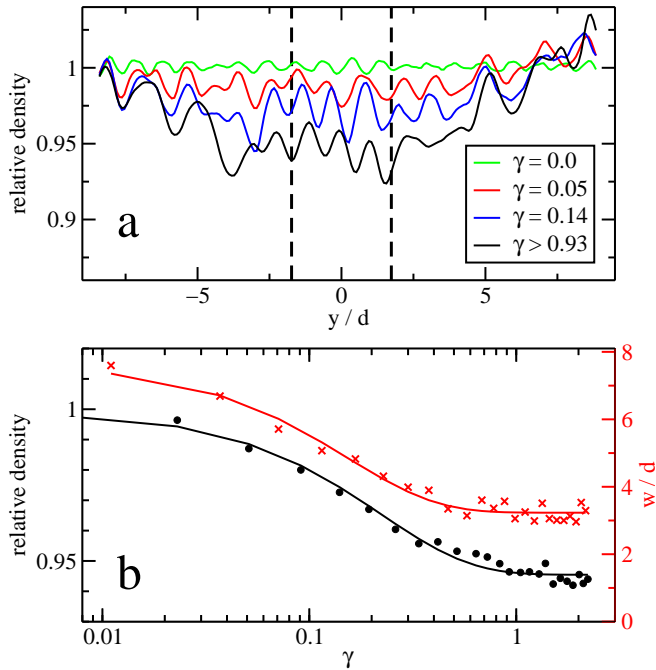}
\caption{(color online). (a) Relative average density as a function of the lateral coordinate $y$ for rape seeds. Data taken from the
same measurement as in Fig.~\ref{width-depth}, the averaging was obtained for the upper part (above $8 d$) of the sample.
The curves correspond to 4 different values of strain $\gamma=0$, $0.05$, $0.14$ and the stationary state
$0.93 < \gamma < 2.22$.
(b) Evolution of the relative density ($\bullet$) and the zone width ($\times$).  The relative density data represent
the average in the middle part of the cell for the range of $-2d <y < 2d$ as indicated with vertical dashed lines
in panel (a).
}
  \label{width-density}
\end{figure}

The transient during which the zone width converges towards its stationary value also involves a shear induced dilation
of the sample. We quantified the evolution of the density by averaging the local intensity of the tomographic image for
the upper part of the sample (height level above $8 d$) over a distance of $x=22 d$ and normalizing 
the data using the first tomogram.  
The resulting lateral density profiles are shown in Fig.~\ref{width-density}(a).
%
%
Here the four curves correspond to $\gamma=0$, $0.05$, $0.14$ and the stationary state $0.93 <\gamma <2.22$.
The central part of the lateral density profiles has been averaged (between the dashed lines in
Fig.~\ref{width-density}(a)) and the resulting data are presented as a function of strain $\gamma$ in
Fig.~\ref{width-density}(b), together with the evolution of the zone width $w$.
As it is seen, the density gradually decreases and in the middle of the sample it reaches about 95 $\%$ of its initial
value. The lines correspond to exponential fits, which give the characteristic strain scales.
They turn out to be somewhat smaller for the evolution of $w$ ($\gamma=0.15$) compared to the case of the density decrease ($\gamma=0.22$).

\subsubsection{X-ray CT measurements in the cylindrical geometry}

The evolution of the system in the cylindrical geometry was tracked using x-ray computed tomography (CT).
We used the robot-based flat panel x-ray C-arm system Siemens Artis zeego of the INKA lab, Otto von Guericke
University, Magdeburg. The recorded volume was 25.2 cm $\times$ 25.2 cm $\times$ 19 cm with a spatial resolution
of 2.03 pixel/mm.
The scanner can record large enough volumes to enable us to use the cylindrical geometry.
The time evolutions of three samples consisting of elongated particles with aspect ratios: $L/d=2.0$, 3.3 and 5.0,
resp., have been measured, by recording about 200 scans in each case. A control measurement with peas has also 
been carried out with a comparable number of scans.

%
%
\begin{figure}[!ht]
\includegraphics[width=\columnwidth]{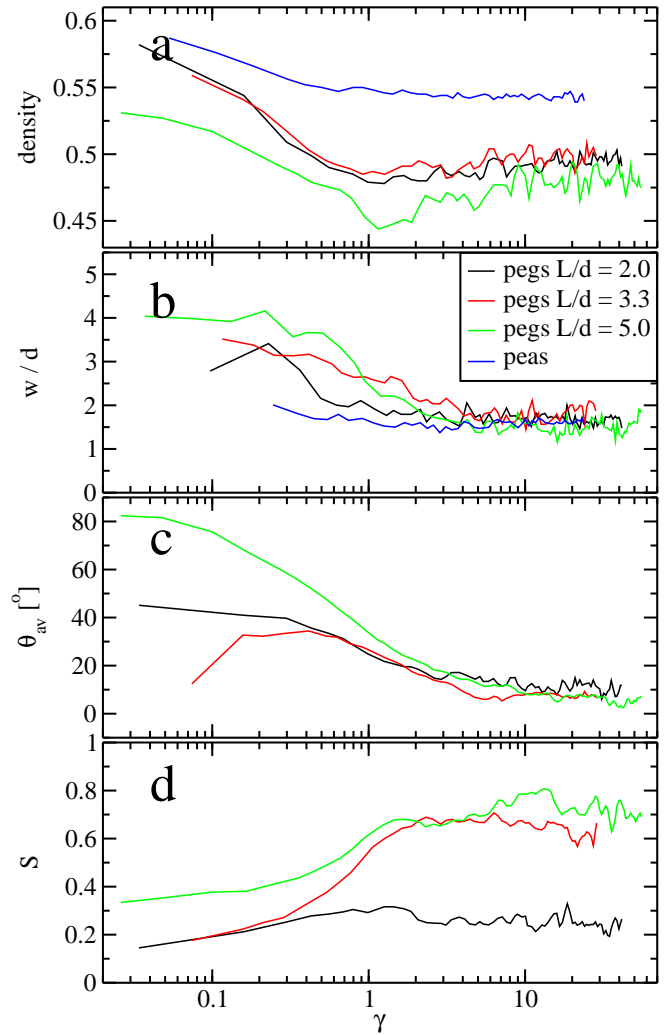}
\caption{(color online). Packing density, zone width $w$, average orientational angle $\Theta_\mathrm{av}$ 
and order parameter $S$ as a function of strain for four materials: cylinders with aspect ratio 
$L/d=2$, 3,3 and 5, resp., and peas.
Measurements were performed with x-ray CT using the cylindrical geometry, with filling height $H/d=11$.
}
  \label{CT-1}
\end{figure}

The advantage of the x-ray CT technique is that it provides detailed information about the
system, similarly to the MRI measurements, but about 10 times faster.
Here we can compare the typical time scales describing the evolution of four important quantities: packing density,
zone width $w$, order parameter $S$, and the average alignment angle $\Theta_\mathrm{av}$.
Here, the order parameter $S$ is defined as the largest eigenvalue of the order tensor, where $S=0$
corresponds to the random isotropic state, while $S=1$ describes the perfectly aligned system \cite{borzsonyi2012}.
The average alignment angle $\Theta_\mathrm{av}$ measures the deviation of the average orientation of the particles with respect to the streamlines.

The density in the shear band rapidly decreases with $\gamma$ (see Fig.~\ref{CT-1}(a)) similarly to the MRI measurements.
The width of the shear zone also decreased, similarly to the other measurements presented above.
The $w(\gamma)$ curves obtained by x-ray CT (Fig.~\ref{CT-1}(b)) are noisier than the corresponding data obtained
by the optical method (Fig.~\ref{time-width-circular}(a)), since the comparably expensive CT technique does 
not allow to collect similar amounts of data as the optical detection (1 vs. 100 measurements, respectively). 
The shear induced ordering is manifested 
by an increasing order parameter $S$ (Fig.~\ref{CT-1}(d)) \cite{borzsonyi2012}, while the average alignment 
angle $\Theta_\mathrm{av}$ decreases and converges towards a stationary value
(Fig.~\ref{CT-1}(c)). 

%
%
\begin{figure}[!ht]
\includegraphics[width=\columnwidth]{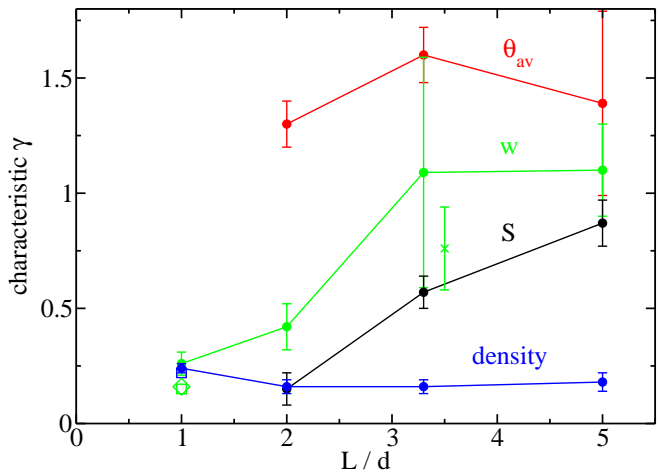}
\caption{(color online). Characteristic strain $\gamma$ describing the evolution of the packing density,
zone width $w$, average
orientational angle $\Theta_\mathrm{av}$ and order parameter $S$ as a function of the aspect ratio $L/d$ of
the particles.  The strain scales corresponding to $w$ and $S$ are both systematically increasing with
$L/d$, while those describing the evolution of the packing density and $\Theta_\mathrm{av}$ are independent of $L/d$.
The majority of the measurements were taken with x-ray CT for peas and pegs ($\bullet$). For comparison, 
data obtained by MRI for rape seeds ($\square$), and by optical detection (surface measurements) for silica 
gel beads ($\Diamond$) and glass rods with $L/d=3.5$ ($\times$) are included.
}
  \label{CT-2}
\end{figure}

The characteristic strain scale $\gamma$ corresponding to the evolution of the above quantities has been
determined by fitting exponential decay functions to these curves. This is shown in Fig.~\ref{CT-2} as a function
of the aspect ratio $L/d$ of the particles. As it is seen, it is the packing density which converges fastest to 
its stationary value, while the slowest process is the evolution of the average alignment angle $\Theta_\mathrm{av}$. 
Both processes are independent of the aspect ratio. The characteristic strain scales needed for the establishment
of the stationary zone width $w$ and the order parameter $S$ both systematically increase with increasing $L/d$.
Thus, the development of shear
induced ordering appears to be correlated with the change in the zone thickness $w$.
Fig.~\ref{CT-2} also includes the zone width data for rape seeds obtained by MRI ($\square$) and for silica gel beads
($\Diamond$) and glass rods with $L/d=3.5$ ($\times$) obtained by optical detection (surface measurements), all of
which are in good agreement with the data obtained by CT.

\section{Numerical model}
\label{num}
\subsection{Description of the fluctuating band model}
\label{numdes}

In order to model the experimental system, numerical simulations have been
performed using the fluctuating band model (see Refs. \cite{torok2007} and
\cite{moosavi2013}). Here we summarize the main ingredients of the
model, which is based on the following assumptions: during quasi static
shear the stress is increased slowly, which first is only accompanied
by elastic deformations. At some point when the system cannot sustain
the stress any more, a plastic event occurs. The system fails along a
path which complies with the boundary conditions and has the smallest
shear force or torque among all possible paths. 

The plastic event rearranges the material and changes its properties
in the vicinity of the yielding path. The accumulated stress is released
and the whole process starts again. The observed velocity profile is
the ensemble average displacement field resulting from these subsequent
yielding events. This is a self-organized process: the
rearrangement along the yielding path (corresponding to the minimal force)
modifies the potential, which is then used to determine the location of
the next plastic event. The above mentioned fundamental features of nonlocality and
self-organization can also be found in other models: e.g. the models
of elastic \cite{jagla2008}, kinetic elasto-plastic
\cite{henann2013,KEP} and shear transformation zone \cite{STZ} theory.

Here, we implement the fluctuating band model in a discretized way
\cite{torok2007,moosavi2013}. Let $a$ be the lengthscale of the coarse
graining \cite{moosavi2013}. We assume perfect translational symmetry
along the shear direction so we project the whole system to a plane
normal to the shear direction. Thus our system is discretized on a
two-dimensional lattice. The  minimal path   is a continuous line initiated
at the bottom at the split and it ends at the top surface.
In the present implementation, we allowed for nearest and next
nearest neighbor connections.

For simplicity, we consider the straight shear cell. The
cylindrical shear cell in this model differs only in a $r^2$ factor
for the torque which is not important for the transient properties.

The shear force can be calculated using Coulomb friction as
\begin{equation}\label{Eq:mindiss}
F=L\int_\text{path} \mueff(y,z) p(z) dl,
\end{equation}
where $L$ is the length of the system in the flow ($x$) direction, the
integral runs along the path from the bottom split to the surface,
$\mueff(y,z)$ is the effective friction coefficient which has spatial
fluctuations, and $p(z)$ is the pressure, which we assume to be the hydrostatic:
$p(z)=\rho g(H-z)$. This is a good approximation far from the walls.
The system is thus characterized by a single scalar
field $\mueff(y,z)$.

The value $\mueff$ is known to have relatively wide distributions,
\cite{Reza08} which are not the same for the preparation and for the
self organized restructuring. Therefore, we define two probability
distributions $P_p(\mueff)$ and $P_r(\mueff)$ for the preparation and
the refresh, respectively. We chose the distributions to be Gaussian
with the restriction $\mueff{\geq}0$:
\begin{equation}\label{Eq:P}
P_i(\mueff)=G_i(m_i,\sigma_i)= C_i
\exp\left(\frac{-(\mueff-m_i)^2}{2\sigma_i^2}\right),
\end{equation}
where $i=\{p,r\}$, and $C_i$ are normalization factors.

Thus in the discretized coarse grained system we have a regular square
lattice of height $\widetilde H=H/a$, where each site is characterized
by a single scalar $\mueff(y,z)$.

We chose the initial distribution of $\mueff$ to match the results of
\cite{Reza08}: $m_p{=}1.5$ and $\sigma_p{=}0.85$. By this choice we
are left with three parameters $\widetilde H$, $m_r$ and $\sigma_r$ to fit the
experimental results.

This simple model describes the material by a scalar variable which
cannot incorporate the orientation of the particles. Therefore, we will restrict the study of this model
at this point to spherical particles and grains with only small shape anisotropy (e.g. sand).

\subsection{Model results}
\label{numres}

%
%
\begin{figure}[!ht]
\includegraphics[width=\columnwidth]{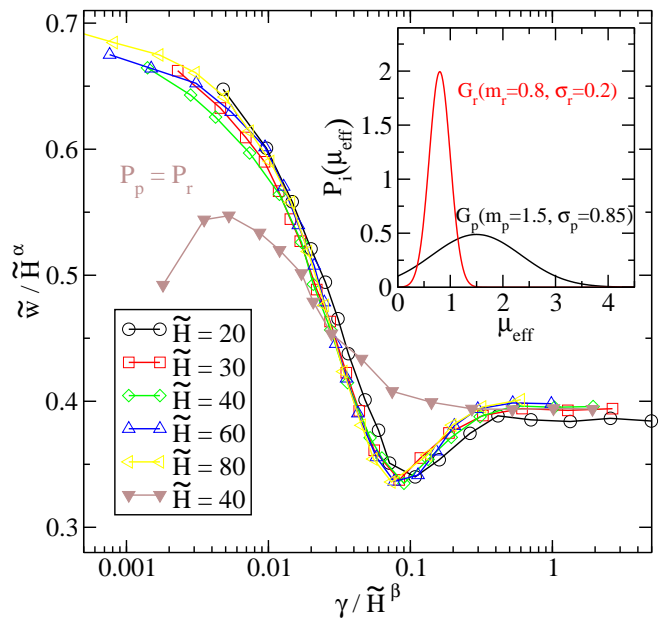}
\caption{(color online). Scaled plot of the time evolution of the width of the shear
zone on the surface of the numerical model system. The preparation and refresh 
probabilities were $P_p=G_p(1.5,0.85)$ and $P_r=G_r(0.8,0.2)$, respectively, which are
illustrated on the inset panel.
The filled triangles show the case where the preparation probability
distribution was the same as the refresh one.}
  \label{Fig:size}
\end{figure}

In the numerical simulations the initially random system is updated 
step by step in order to mimic the experimental procedure.
At each wall displacement step the minimization gives us an
instantaneous path which divides the system into two moving
blocks. The displacement at each position was recorded for every step
and averaged for many independent realizations. The resulting
surface displacement fields were fitted the same way as in the
experiments, from which the transient evolution of the reduced shear
zone width $\widetilde w=w/a$ can be calculated.

The numerical procedure involves a length factor $\widetilde H=H/a$.
In Fig.~\ref{Fig:size} we show the results of several runs using
the same $P_r=G_r(0.8,0.2)$ but different $\widetilde H$.
The curves are collapsed on top of each other with the following
scaling relations:

\begin{align}
w^* &= \widetilde w/\widetilde H^\alpha\cr
\gamma^* &= \gamma/\widetilde H^\beta.
\end{align}
The scaling exponents were found to be $\alpha=0.6$ and
$\beta=1.0$.
The first exponent is in agreement with the results of Jagla \cite{jagla2008}
obtained for the stationary zone width.
Thus, in our model the functional form of the curve describing the
evolution of the shear zone width in scaled variables is independent
of the length factor.

This result has two important implications: (i) in order to fit the
functional form of the evolution of the width of the shear zone we
have only two independent parameters left ($m_r$, $\sigma_r$), (ii) once the
functional form is found, the length factor $\widetilde H$ and the coarse
graining factor $a$ can be determined: converting back to unscaled
variables the width of the shear zone is $\widetilde w a$, and the height of
the system is $\widetilde H a$. The values of $a$ and $\widetilde H$
are determined using the stationary values of $w$ and $H$ from the experiments.

In Fig.~\ref{Fig:size} we have included another curve in which we
changed the preparation distribution $P_p$ of $\mueff$ to make it coincide to 
the refresh distribution $P_r$. As expected, the stationary state has the
same width as in the previous case but the transient is different. The
width of the shear zone at the beginning is larger than in the
stationary state but the local minimum is not present.
This result is independent of the actual value of the distribution, if
$P_p=P_r$ we always see the same behavior.

The fact that the width of the shear zone decreases can be best
understood through the one-dimensional version of the fluctuating band
model which is the Bak-Sneppen model of evolution \cite{BakSneppen}.
In that model, a one dimensional lattice is filled with uniform random
numbers between 0 and 1. The smallest is chosen along with its two
neighbors, and they are refreshed from the same random distribution.
The stationary state is characterized by avalanche like dynamics of
the minimal site, and the inactive sites have values distributed
between 0.66 and 1. Thus even if the refreshing distribution is the
same as the initial one, the resulting potential will be different from
the original and it will be composed of larger values than the mean of
the refreshing distribution.

This is also the case here. The average values of $\mueff$ in the
stationary state in the shear zone can be measured. Using the previous
examples for the case of $P_r=G_r(0.8,0.2)$, the stationary average value
is $\langle\mueff\rangle =0.83$ instead of $\sim0.8$ with standard
deviation of $0.058$ instead of $0.2$. So after the
initial transient the actual field of $\mueff(y,z)$ has larger values
with narrower distribution than the mean of $P_r$, which pushes the
minimal paths towards shorter length. This narrows the zone,
and explains the initial decrease of the width of the shear
zone.

%
%
\begin{figure}[!ht]
\includegraphics[width=\columnwidth]{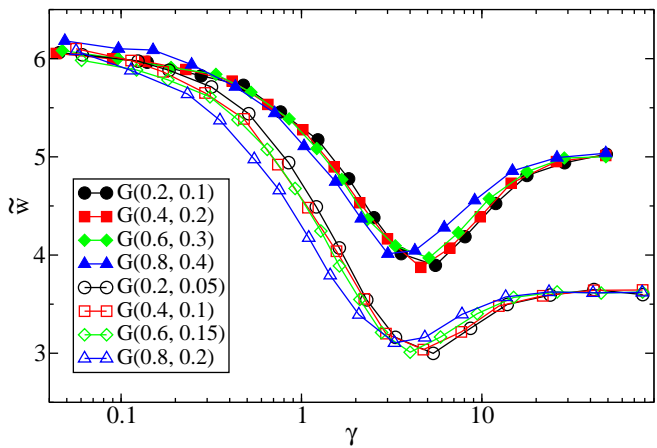}
\caption{(color online). Evolution of surface width obtained by numerical simulations
for different
parameters $m_r$ and $\sigma_r$ of the refresh probability distribution.
For filled symbols $\sigma_r/m_r=0.5$ while for open symbols
$\sigma_r/m_r=0.25$. The preparation distribution was $P_p=G_p(1.5,0.85)$.
}
  \label{Fig:sigmad} 
\end{figure}

In Fig.~\ref{Fig:sigmad}, two sets of curves are shown for different
combinations of  $m_r$ and $\sigma_r$. All of them are for the same system size.
The curves fall on
top of each other (apart from a tiny shift in shear strain) if $\sigma_r/m_r$
is kept constant. This is a surprising result for which we have no
theoretical explanation. Thus for fitting the scaled $w/w_\text{stationary}$ 
curves we are left with only one parameter:
$\sigma_r/m_r$.

%
%
\begin{figure}[!ht]
\includegraphics[width=\columnwidth]{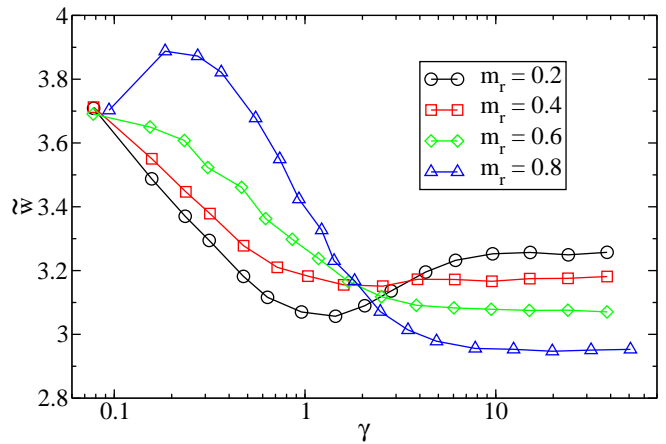}
\caption{(color online). Evolution of surface width obtained by numerical simulations 
for different $m_r$, with $P_p=G_p(0.8,0.7)$ and $P_r=G_r(m_r,0.7)$.
}
  \label{Fig:change_d}
\end{figure}

Now let us consider the minimum of the zone width during its evolution.
In Fig.~\ref{Fig:change_d} we plot
several curves showing the evolution of the zone width, where
only $m_r$ is varied. As
mentioned before, if $m_r=m_p$ there is no minimum, but as $m_r$
decreases the minimum appears and becomes more and more pronounced.
This happens because after the initial decrease of the width, many contacts
remain with high effective friction coefficients (originating from the
initial preparation), enforcing a narrow zone.
After the minimum of the width is reached
the minimal path continues to fluctuate and gradually erases the
preparation state. This can be seen in Fig.~\ref{Fig:change_d}, as the
stationary state is reached at the same point at around $\gamma\simeq8$  

%
%
\begin{figure}[!ht]
\includegraphics[width=0.9\columnwidth]{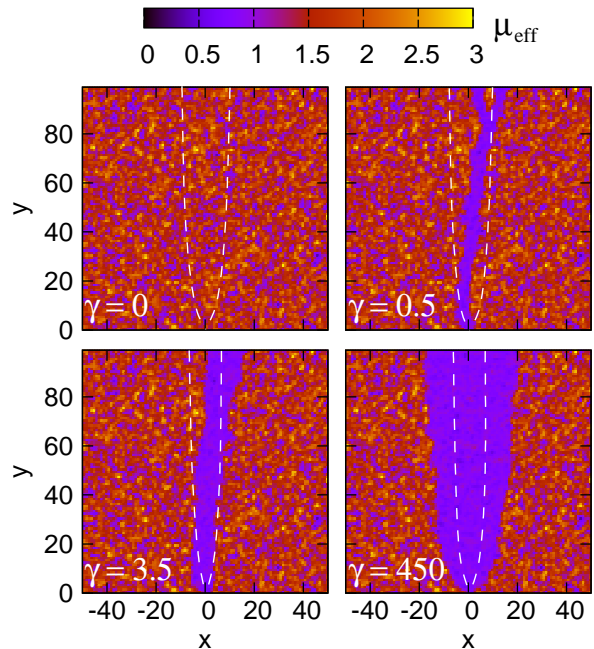}
\caption{(color online). Snapshots of $\mu_\text{eff}$ in the numerical model system 
at $\gamma$ values of 0, 0.5, 2.5, 450. The height of the system was $H=100$, 
and the probability distributions were $P_p=G_p(1.5,0.85)$ and 
$P_r=G_r(0.8,0.2)$, as for the dataset presented with $\triangle$ in 
Fig.~\ref{Fig:sigmad}. 
The white dashed lines show the actual width of the shear zone.
}
  \label{Fig:rho_t}
\end{figure}

This scenario can be followed on a selected example
shown in Fig.~\ref{Fig:rho_t} which was created using
$P_p=G_p(1.5,0.85)$ and $P_r=G_r(0.8,0.2)$. The dotted lines show the
shape of the shear zone in the stationary state. The first snapshot 
corresponds to the preparation state. One can already spot the blue patch at the top
of the sample at around $y{=}12$ where the first path will be formed
(right branch on the second snapshot), which is relatively far from the
center. This type of fluctuations account for the large initial width 
of the zone as calculated from the ensemble of many realizations. 
The path then moves inwards as seen on the third snapshot corresponding 
to the moment when $w(\gamma)$ takes its minimum. 
The last panel shows the stationary state where the preparation state 
has been erased in a wider region. It is also interesting that the width 
of the shear zone is much smaller than the actual space explored by the
fluctuating path. This fact was also observed in a previous study \cite{sakaie2008}.

The experimental data were fitted taking into account the above
described features of the model. This approach fits only the data for
the silica gel beads reasonably, as shown in Fig.~\ref{time-width-circular} (a).
The parameters are $a=3.1$, $\sigma_r/m_r=0.38$.
For corundum, the initial decrease of the width could be
recovered but not the slow evolution of the minimum. The model has
only one internal strain scale, while real systems appear to have two:
one for the initial decrease of the zone width and another one for the
later evolution, when the fluctuating path erases the preparation
state. This can be easily incorporated into the model by adding
a new strain scale $\gamma_0$ in the following way. At each step the
local effective friction coefficient is generated 
(for each (y,z) site) using a probability:
\begin{equation}\label{Eq:Ptau}
	P_r(\gamma)=G_r +(G_p-G_r)e^{\gamma/\gamma_0}
\end{equation}
 
This means, that instead of an abrupt change we allow for an exponential
decay from the preparation to the refresh distribution depending on the local
shear strain. This way, the experimentally observed evolution of the
zone width for corundum can be reproduced using the exponential constant
of $\gamma_0=1$, which is illustrated on Fig.~\ref{time-width-circular} (b).
The other parameters are: $\sigma_r/m_r=0.33$ and $a=1.95$.

The resulting $\sigma_r/m_r$ factor is close to the the value obtained
with the silica gel beads, the difference is only 15\%. The coarse-graining
length is larger by 50\% for beads than for irregular shaped corundum,
which is in accordance with other observations \cite{Pena07} that
smoother grains have larger length scales in the system.

To summarize the numerical results, we have shown that the fluctuating
band model has only two important independent parameters from which the
coarse-graining length $a$ can be fitted by the stationary width of the
shear zone. The second parameter $\sigma_r/m_r$ accounts for the transient
behavior of the model. The initial decrease of the width of the shear
zone is the result of the shear band removing the easily deformable
parts of the system. The refreshing-probability distribution for the
effective friction coefficient is narrower than that of the initially 
prepared ensemble,
resulting in a minimum in the zone width due to remaining
large values of the preparation effective friction coefficients. 
This is later erased by the fluctuating path. The results show that the
refreshing probability appears to depend locally on the
shear strain. It introduces another time scale that describes the
evolution of the zone in the second stage.

\section{Summary}

The evolution of the shear zone was investigated experimentally using surface (optical) and bulk (x-ray CT and MRI)
measurements for various dry granular material placed in straight and cylindrical split bottom shear cells.
The behavior of materials consisting of beads, irregular grains (e.g. sand) and elongated particles have been compared.
The experimental findings have been fitted with the results of numerical simulations based on a fluctuating band model.
When an initially random sample is sheared the width of the shear zone
significantly decreases in the first stage of the process. The characteristic strain associated with
this decrease is about $\gamma=1$, and is systematically increasing with increasing shape anisotropy, i.e.
when the grain shape changes from spherical to irregular (e.g. sand) and becomes elongated (pegs).
For rods, both the characteristic strain necessary for the evolution of the zone width
and the shear induced order parameter increase with increasing particle aspect ratio $L/d$, while the
characteristic strain corresponding to the evolution of packing density and shear alignment direction
appears to be independent of $L/d$.
The shrinking of the shear zone observed in the first stage of the process is followed by a slight widening, 
which is more pronounced for
rod like particles than for grains with smaller shape anisotropy (spherical beads or irregular particles).
The final zone width is significantly smaller for irregular grains when compared to the case of spherical beads.

The results obtained for spherical particles and grains with only small shape 
anisotropy (e.g. sand) can be recovered using the
fluctuating band model with only two parameters: the ratio of the
two variables $\sigma_r/m_r$ of the distribution of the effective friction coefficient in
the stationary state and a strain scale $\gamma_0$. The initial evolution of the width of
the shear zone in this numerical model is a result of self-organized
modifications of the local effective friction coefficient by the
fluctuating path which may favor narrower or wider shear zones.

\section{Acknowledgements}
Financial support by the DAAD/M\"OB researcher exchange program (grant no. 29480), the Hungarian Scientific
Research Fund (grant no. OTKA NN 107737), and the J\'anos Bolyai Research Scholarship of the Hungarian Academy
of Sciences is acknowledged.


\begin{thebibliography}{99}

\bibitem{losert2000}
W. Losert, L. Bocquet, T.C. Lubensky, and J.P. Gollub,
Phys. Rev. Lett. {\bf 85}, 1428 (2000).

\bibitem{veje1999}
C. T. Veje, D. W. Howell, and R. P. Behringer,
Phys. Rev. E {\bf 59}, 739 (1999͒).

\bibitem{shojaaee2012}
Z. Shojaaee, J-N. Roux, F. Chevoir and D.E. Wolf,
Phys. Rev. E {\bf 86}, 011301 (2012).

\bibitem{mueth2000}
D.M. Mueth, G.F. Debregeas, G.S. Karczmar, P.J. Eng, S.R. Nagel, and H.M. Jaeger,
Nature {\bf 406} 385 (2000).

\bibitem{tsai2004}
J.-C. Tsai and J.P. Gollub,
Phys. Rev. E {\bf 70}, 031303 (2004).

\bibitem{wegner2014}
S. Wegner, R. Stannarius, A. Boese, G. Rose, B. Szab\'o, E. Somfai, and
T. B\"orzs\"onyi,
Soft Matter, {\bf 10}, 5157 (2014).

\bibitem{dijksman2010}
J.A. Dijksman and M. van Hecke,
Soft. Matt. {\bf 6}, 2901 (2010).

\bibitem{fenistein2003}
D. Fenistein and M. van Hecke,
Nature {\bf 425}, 256 (2003).

\bibitem{fenistein2004}
D. Fenistein, J.W. van de Meent and M. van Hecke,
Phys. Rev. Lett. {\bf 92}, 094301 (2004).

\bibitem{fenistein2006}
D. Fenistein, J.W. van de Meent and M. van Hecke,
Phys. Rev. Lett. {\bf 96}, 118001 (2006).

\bibitem{cheng2006}
X. Cheng, J.B. Lechman, A. Fernandez-Barbero, G.S. Grest, H.M. Jaeger,
G.S. Karczmar, M.E. M\"obius, and S.R. Nagel
Phys. Rev. Lett {\bf 96}, 038001 (2006).

\bibitem{nichol2012}
K. Nichol and M. van Hecke,
Phys. Rev. E {\bf 85}, 061309 (2012).

\bibitem{wortel2014}
G.H. Wortel, J.A. Dijksman and M. van Hecke,
Phys. Rev. E {\bf 89}, 012202 (2014).

\bibitem{henann2013}
D.L. Henann and K. Kamrin,
PNAS {\bf 110}, 6730 (2013).

\bibitem{unger2004}
T. Unger, J. T\"or\"ok, J. Kert\'esz, and D.E. Wolf,
Phys. Rev. Lett. {\bf 92}, 214301 (2004).

\bibitem{torok2007}
J. T\"or\"ok, T. Unger, J. Kert\'esz, and D.E. Wolf,
Phys. Rev. E {\bf 75}, 011305 (2007).

\bibitem{moosavi2013}
R. Moosavi, M.R. Shaebani, M. Maleki, J. T\"or\"ok, D.E. Wolf, and W. Losert,
Phys. Rev. Lett. {\bf 111}, 148301 (2013).

\bibitem{depken2006}
M. Depken, W. van Saarloos and M. van Hecke,
Phys. Rev. E {\bf 73}, 031302 (2006).

\bibitem{depken2007}
M. Depken, J.B. Lechman, M. van Hecke, W. van Saarloos and G. S. Grest
EPL {\bf 78}, 58001 (2007).

\bibitem{sakaie2008}
K. Sakaie, D. Fenistein, T.J. Caroll, M. van Hecke, and P. Umbanhowar,
EPL  {\bf 84}, 38001 (2008).

\bibitem{luding2008}
S. Luding,
Particuology {\bf 6}, 501-505 (2008).

\bibitem{ries2007}
A. Ries, D.E. Wolf, and T. Unger,
Phys. Rev. E {\bf 76}, 051301 (2007).

\bibitem{azema2013}
E. Az\'ema, F. Radja\"i, B. Saint-Cyr, J.-Y. Delenne, and P. Sornay
Phys. Rev. E {\bf 87}, 052205 (2013).

\bibitem{utter2004}
B. Utter and R.P. Behringer,
Eur. Phys. J. E {\bf 14}, 373-380 (2004).

\bibitem{farhadi2014}
S. Farhadi and R. P. Behringer,
Phys. Rev. Lett., {\bf 112}, 148301 (2014).

\bibitem{kabla2009}
A. J. Kabla and T. J. Senden,
Phys. Rev. Lett., {\bf 102}, 228301 (2009).


\bibitem{toiya2004}
M. Toiya, J. Stambaugh and W. Losert,
Phys. Rev. Lett. {\bf 93}, 088001 (2004).

\bibitem{jagla2008}
E.A. Jagla,
Phys. Rev. E {\bf 78}, 026105 (2008).

\bibitem{KEP}
V. Mansard, A. Colin, P. Chauduri, and L. Bocquet,
Soft Matter {\bf 7}, 5524-5527 (2011).

\bibitem{STZ}
M. L. Falk, J. S. Langer, and L. Pechenik
Phys. Rev. E {\bf 70}, 011507 (2004).

\bibitem{Reza08}
M. R. Shaebani, T. Unger, J. Kert\'esz,
Phys. Rev. E {\bf 78}, 011308 (2008).

\bibitem{borzsonyi2012}
T. B\"orzs\"onyi, B. Szab\'o, G. T\"or\"os, S. Wegner, J. T\"or\"ok, E. Somfai, T. Bien, and R. Stannarius,
Phys. Rev. Lett. {\bf 108}, 228302 (2012).

\bibitem{wegner2012}
S. Wegner, T. B\"orzs\"onyi, T. Bien, G. Rose, and R. Stannarius,
Soft Matter {\bf 8}, 10950 (2012).

\bibitem{borzsonyi2012-2}
T. B\"orzs\"onyi, B. Szab\'o, S. Wegner, K. Harth, J. T\"or\"ok, E. Somfai, T. Bien, and R. Stannarius,
Phys. Rev. E {\bf 86}, 051304 (2012).

\bibitem{singh2014}
A. Singh, V. Magnanimo, K. Saitoh, and S. Luding,
arXiv, 1312.7133 (2014).

\bibitem{BakSneppen} P. Bak, K. Sneppen, 
Phys. Rev. Lett. {\bf 71}, 4083 (1993).

\bibitem{Pena07} A.A. Pe\~na, R. Garc\'{\i}a-Rojo, H. J. Herrmann,
Granular Matter {\bf 9}, 279-291 (2007).

\end{thebibliography}
\end{document}